\documentclass[a4paper,12pt,onecolumn]{article}
\usepackage{amsmath,amsthm,amssymb}
\usepackage{mathtext}
\usepackage[T1,T2A]{fontenc}
\usepackage[utf8]{inputenc}
\usepackage[english]{babel}
\usepackage{graphicx}
\usepackage{amssymb}
\usepackage{amsmath}
\usepackage{setspace}
\usepackage{cite}

\begin{document}

\begin{center}
	{\large
		{\bf 
			Monte Carlo simulation of mixed spin-orbit ordering
		} 
	}
	\vskip0.5\baselineskip{
		\bf 
		Yu. D. Panov$^1$, V.A. Ulitko$^1$
	}
	\vskip0.5\baselineskip{
		$^{1}$Ural Federal University, 620002 Ekaterinburg, 19 Mira street, Russia
	}
	
\end{center}

\begin{abstract}
The effect of a magnetic field on the thermodynamic properties of the Kugel–Khomskii model with maximal entanglement using the quasi‑classical Monte Carlo method with the heat bath algorithm. The temperature dependencies of single‑site correlators show that the application of a magnetic field results in non-zero magnetization and the simultaneous appearance of an orbital ordering at the same point in the Brillouin zone where a non-zero spin-orbit correlator exists.
\end{abstract}


\section*{INTRODUCTION}
Models that take into account the orbital contribution to the exchange interaction are widely used to describe such strongly correlated compounds as cuprates (KCuF$_{3}$, K$_{2}$CuF$_{4}$), manganites (RMnO$_{3}$, La$_{1-x}$Sr$_{1+x}$MnO$_{4}$), vanadates (V$_{2}$O$_{3}$, YVO$_{3}$, LaVO$_{3}$), ruthenates (Sr$_{2}$RuO$_{4}$, Ca$_{2}$RuO$_{4}$), chromites (CdCr$_{2}$O$_{4}$, HgCr$_{2}$O$_{4}$), rare earth magnets (Tb$_{2}$Ti$_{2}$O$_{7}$, Tb$_{2}$Sn$_{2}$O$_{7}$) and many others, including new "<breathing"> and cluster magnets (Ba$_{3}$Yb$_{2}$Zn$_{5}$O$_{11}$, LiZn$_{2}$Mo$_{3}$O$_{8}$, $\kappa$-(BEDT-TTF)$_{2}$Cu$_{2}$(CN)$_{3}$) \cite{Kugel1982,Tokura2000,Leonov2022,Chen2024}. These systems implement a wide range of phase states, including various types of magnetic and orbital ordering.

One of the simplest models of spin‑orbital superexchange is the Kugel--Khomskii model \cite{Kugel1982}, however, taking into account the explicit composition of the ground-state multiplet leads to various modifications of the Hamiltonian (see, e.g., \cite{Chizhov2026}). The analytical solution of the Kugel–Khomskii models is difficult due to the large number of degrees of freedom, the competition of various types of ordering, and the complexity of accounting for correlations between spin and orbital subsystems. 

A remarkable feature of Kugel--Khomskii models is quantum entanglement, which gives rise to the possibility of non-trivial ordered states that cannot be reduced to a spin and/or orbital ordering. In this article, we explore the effect of a magnetic field on the thermodynamic properties of the Kugel-–Khomskii model with maximal entanglement \cite{Valiulin2020}. For this purpose, we use the quasi‑classical Monte Carlo (MC) method with the heat bath algorithm, which we previously developed to study the phase diagrams of multi-orbital models of cuprates and nickelates \cite{Panov2025,Ryumshin2025}. We calculate the temperature dependencies of energy, heat capacity, as well as single‑site spin, orbital, and spin-orbital correlators. We show that a magnetic field results in non-zero magnetization and the simultaneous appearance of an orbital correlator at the same point in the Brillouin zone where a non-zero spin-orbit correlator exists.

\section*{The heat bath algorithm for Kugel--Khomskii model}

An exact diagonalization study of the ground state of a finite chain for Kugel-Khomskii models with various symmetries \cite{Valiulin2020} shows that strong quantum entanglement between the orbital and spin subsystems occurs in the model with the Hamiltonian
\begin{equation}
	{\hat{H}} = J \sum_{\left\langle ij \right\rangle } ({\hat{\boldsymbol{{S}}}}_{i} , {\hat{\boldsymbol{{S}}}}_{j}) {\hat{{T}}}_{z,i} {\hat{{T}}}_{z,j} - h \sum_{i} {\hat{{S}}}_{z,i} .
	\label{eq:Hamiltonian}
\end{equation}
Here, $\hat{S}_{\alpha,i}$ are spin operators, and $\hat{T}_{\alpha,i}$ are pseudospin ones, related to orbital degrees of freedom, and $\alpha$ is a spin and pseudospin component index. We explore the model \eqref{eq:Hamiltonian} on a square lattice. In the first term, the summation is performed over nearest neighbors. We consider the case $S = 1/2$, $T = 1/2$, and, subsequently, incorporate the values of $S$ and $T$ into the constants $K$ and $h$.

Let's briefly formulate the heat bath algorithm \cite{Miyatake1986} for the model \eqref{eq:Hamiltonian}. We write the system's wave function as a product of wave functions at the lattice sites:
\begin{equation}
	\left| \Psi \right\rangle = \prod_{i} \left| \psi_{i} \right\rangle , \;\; \left| \psi_{i} \right\rangle = \sum_{\mu,\tau} c_{\mu\tau}^{(i)} \left| \mu \right\rangle \left| \tau \right\rangle .
	\label{eq:Psi}
\end{equation}
Here, $\left| \mu \right\rangle$ and $\left| \tau \right\rangle$ runs over the basis spin-up and spin-down functions (eigenfunctions of the $\hat{S}_{z}$ and $\hat{T}_{z}$ operators). We find the Hamiltonian for site $i$ by averaging over the states of all sites $j \neq i$:
\begin{equation}
	\hat{H}_{i} = \big\langle \tilde{ \Psi }_{i} \big| \hat{H} \big| \tilde{ \Psi }_{i} \big\rangle , \quad \big| \tilde{ \Psi }_{i} \big\rangle = \prod_{j \neq i} \left| \psi_{j} \right\rangle.
\end{equation}
Up to a constant term, we obtain 
\begin{equation}
	\hat{H}_{i} = J \sum_{ i' } \big( \big\langle \hat{\boldsymbol{S}} \big\rangle_{i'} , \hat{\boldsymbol{S}} \big) \big\langle \hat{T}_{z} \big\rangle_{i'} \hat{T}_{z} - h \hat{S}_{z} .
	\label{eq:Hc}
\end{equation}
Here, $\big\langle \hat{A} \big\rangle_{i'} = \big\langle \psi_{i'} \big| \hat{A} \big| \psi_{i'} \big\rangle$ are the molecular fields acting on site $i$ from site $i'$. Summation is performed over the nearest neighbors of site $i$. We change the site state using the heat bath algorithm. Solving the eigenvalue problem at site $i$,
\begin{equation}
	\hat{H}_{i} \left| \psi_{i,\nu} \right\rangle = \varepsilon_{\nu} \left| \psi_{i,\nu} \right\rangle , \quad \nu = 1, \ldots 4 ,
\end{equation}
we construct the Boltzmann's distribution function:
\begin{equation}
	F(n) = \sum_{\nu=1}^{n} p_{\nu} , \quad p_{\nu} = \frac{e^{- \beta\varepsilon_{\nu} }}{\sum_{\mu} e^{- \beta \varepsilon_{\mu} }} .
\end{equation}
Let $\gamma$ be a random variable uniformly distributed over a unit interval. From the inequality $F(n-1) < \gamma \leqslant F(n)$ we find the number $n$ and define the new state for site $i$ as $\left| \psi_{i,n} \right\rangle$. Next, we calculate the required quantum mechanical averages and collect statistics on the MC steps in a standard manner, using the wave function \eqref{eq:Psi} for the current state of the system.

\section*{RESULTS}

The simulation was performed on a 32$\times$32 square lattice with periodic boundary conditions. A parallel version of the program for graphics accelerators was implemented using CUDA SDK tools. The cuSOLVER library was used to diagonalize the matrices. Averaging was performed over $10^{5}$ MC steps per node after $10^{5}$ MC steps for annealing at each temperature. The temperature step was $\Delta T / |J| = 0.05$.

We calculated the temperature dependencies of the internal energy and specific heat, as well as the average values of single‑site correlators at the $\Gamma$ and $M$ points of the Brillouin zone ($\boldsymbol{k} = (0,0)$ and $\boldsymbol{k} = (\pi,\pi)$).
\begin{equation}
	A_{\boldsymbol{k}} = \overline{ \Big | 	\frac{1}{N} \sum_{j} e^{i(\boldsymbol{k},\boldsymbol{r}_{j})} \big\langle \hat{A} \big\rangle_{j} \Big| } .
	\label{eq:corr}
\end{equation}
Here, $N$ is the number of lattice sites, and the overline indicates averaging over MC steps. The operator $\hat{A}$ is equal to $\hat{\boldsymbol{S}}$ for the spin correlator, $\hat{\boldsymbol{T}}$ for the pseudospin one, and $\hat{\boldsymbol{S}} \hat{T}_{z}$ for the spin‑orbital one.

\begin{figure}
	\centering
	\includegraphics[width=\linewidth]{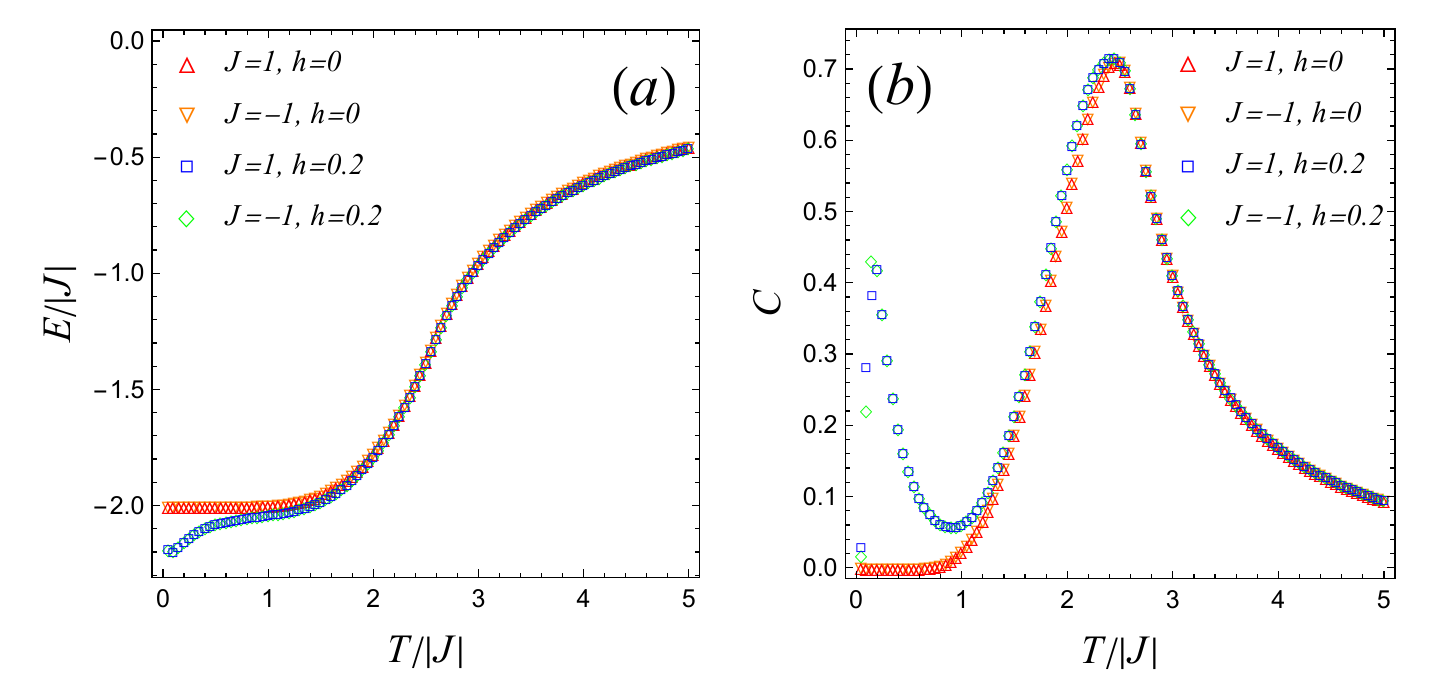}
	\caption{(a) The energy and (b) the specific heat for $J=\pm1$, $h=0.0$ and $h=0.2$.}
	\label{fig:EnC}
\end{figure}

Figure \ref{fig:EnC} shows the temperature dependencies of energy and specific heat for $J=\pm1$, $h=0.0$ and $h=0.2$. The value of $h$ is chosen to be comparable to and smaller than $|J|$ in order to illustrate the effect of an external field on the ordered state emerging in the system. For energy, the dependencies at $J=\pm1$ practically coincide both in the zero field and in the non‑zero field. 

\begin{figure}
	\centering
	\includegraphics[width=0.9\linewidth]{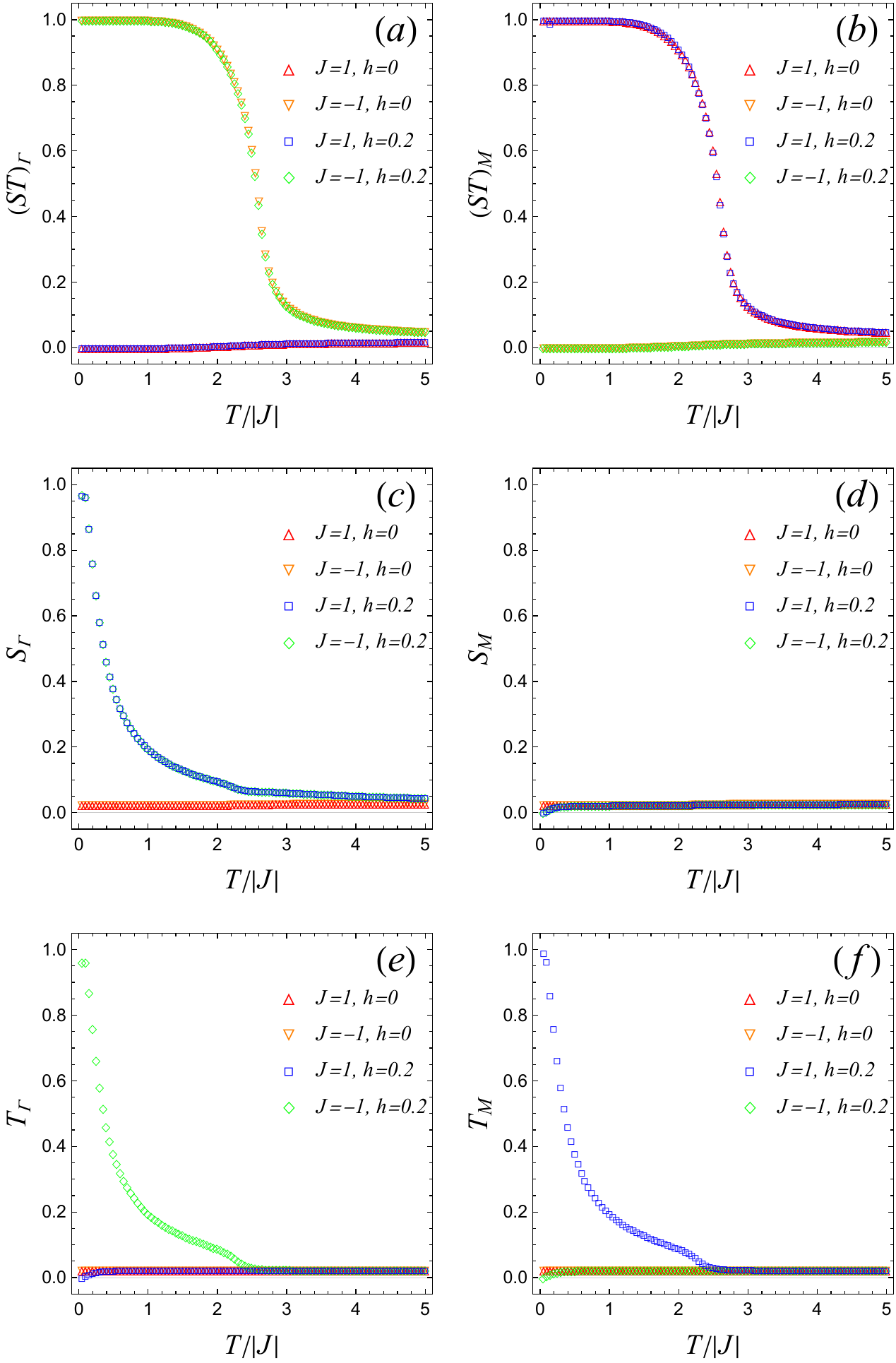}
	\caption{The single-site correlators, Eq.\eqref{eq:corr}, for  $J=\pm1$, $h=0.0$ and $h=0.2$: 	spin-orbital correlator at the (a) $\Gamma$ point, (b) $M$ point; spin correlator at the (c) $\Gamma$ point, (d) $M$ point; orbital correlator at the (e) $\Gamma$ point, (f) $M$ point.
	}
	\label{fig:corr}
\end{figure}

Figure \ref{fig:corr} shows the temperature dependences of single-site correlators at the $\Gamma$ and $M$ points of the Brillouin zone at $J=\pm1$, $h=0.0$ and $h=0.2$. It can be seen that the high-temperature heat-capacity peak at $T/|J|\simeq2.4$ is associated with a specific spin-orbital ordering described by the correlator $(ST)_{\Gamma}$ for $J=-1$ (Fig.\ref{fig:corr}a) and $(ST)_{M}$ for $J=1$ (Fig.\ref{fig:corr}b). The values of $(ST)_{\boldsymbol{k}}$ are practically identical for $h=0.0$ and $h=0.2$. In a zero field, the correlators $S_{\boldsymbol{k}}$ (Fig.\ref{fig:corr}c,d) and $T_{\boldsymbol{k}}$ (Fig.\ref{fig:corr}e,f) remain zero down to the lowest temperatures. The low-temperature heat capacity peak at $T/|J| \simeq 0.1$ is associated with the emergence of magnetization in a non-zero magnetic field. Orbital ordering is induced simultaneously: for $J=-1$, the correlator $T_{\Gamma}$ becomes non-zero, whereas for $J=1$, $T_{M} \neq 0$ appears. The correlator $S_{M}$ remains zero in all cases.

\section*{СONCLUSION}

We explore the effect of a magnetic field on the thermodynamic properties of the Kugel–Khomskii model with maximal entanglement \citen{Valiulin2020} using the quasi‑classical Monte Carlo method with the heat bath algorithm. The temperature dependencies of energy, heat capacity, and single‑site spin, orbital, and spin-orbital correlators for the $\Gamma$ and $M$ points of the Brillouin zone have been calculated. We find that in the zero field, only the corresponding spin-orbital correlator is non-zero. The application of a magnetic field results in non-zero magnetization and the simultaneous appearance of an orbital correlator at the same point in the Brillouin zone where a non-zero spin-orbit correlator exists.

It should be noted that the obtained results pertain to the improved mean-field description provided by the heat bath algorithm. A more accurate treatment of the quantum fluctuations associated with quantum entanglement requires, at the very least, going beyond the semiclassical approximation \eqref{eq:Psi}. Generalizing the approach used to clusters of multiple nodes and partially or fully accounting for quantum correlations within the cluster are subjects for further research.

\subsection*{FUNDING}
The work was supported by the Ministry of Science and Higher Education of the Russian Federation, the project FEUZ-2026-0014.

\subsection*{CONFLICT OF INTEREST}
The authors declare that they have no conflicts of interest. 


\bibliographystyle{elsarticle-num} 

\bibliography{MSN2026}

@article{Kugel1982,
author = {Kugel', Kliment I and Khomskiĭ, D I},
doi = {10.1070/PU1982v025n04ABEH004537},
issn = {0038-5670},
journal = {Soviet Physics Uspekhi},
month = {apr},
number = {4},
pages = {231--256},
title = {{The Jahn-Teller effect and magnetism: transition metal compounds}},
volume = {25},
year = {1982}
}

@article{Tokura2000,
author = {Tokura, Y. and Nagaosa, N.},
doi = {10.1126/science.288.5465.462},
issn = {0036-8075},
journal = {Science},
month = {apr},
number = {5465},
pages = {462--468},
pmid = {10775098},
title = {{Orbital Physics in Transition-Metal Oxides}},
volume = {288},
year = {2000}
}

@article{Chen2024,
author = {Chen, Gang V. and Wu, Congjun},
doi = {10.1038/s41535-023-00614-2},
isbn = {4153502300614},
issn = {2397-4648},
journal = {npj Quantum Materials},
month = {jan},
number = {1},
pages = {1},
publisher = {Springer US},
title = {{Multiflavor Mott insulators in quantum materials and ultracold atoms}},
volume = {9},
year = {2024}
}

@article{Leonov2022,
    title = {{Charge-orbital ordering, magnetic state, and exchange couplings in quasi-one-dimensional V$_{6}$O$_{13}$}},
    author = {Leonov, I. V.},
    journal = {Письма в ЖЭТФ},
    volume = {116},
    issue = {12},
    pages = {853},
    year = {2022},
    doi = {10.31857/S1234567822240065}
}

@article{Chizhov2026,
author = {Chizhov, D.E. and Igoshev, P.A. and Irkhin, V.Yu.},
doi = {10.1016/j.physleta.2026.131481},
issn = {03759601},
journal = {Physics Letters A},
month = {may},
pages = {131481},
title = {{Quantum magnetic phase transitions in a Kugel-Khomskii model including spin-orbit coupling}},
volume = {579},
year = {2026}
}

@article{Valiulin2020,
author = {Valiulin, V. E. and Mikheyenkov, A. V. and Chtchelkatchev, N. M. and Kugel, K. I.},
doi = {10.1103/PhysRevB.102.155125},
issn = {2469-9950},
journal = {Physical Review B},
month = {oct},
number = {15},
pages = {155125},
publisher = {American Physical Society},
title = {{Quantum entanglement, local indicators, and the effect of external fields in the Kugel-Khomskii model}},
volume = {102},
year = {2020}
}

@article{Panov2025,
author = {Panov, Yu D. and Ulitko, V. A. and Yasinskaya, D. N. and Moskvin, A. S.},
doi = {10.1134/S0031918X25600976},
issn = {15556190},
journal = {Physics of Metals and Metallography},
number = {10},
pages = {1216--1221},
title = {{Modified Monte Carlo Method with the Heat Bath Algorithm for a Model Cuprate}},
volume = {126},
year = {2025}
}

@article{Ryumshin2025,
author = {Ryumshin, V.S. and Panov, Yu.D. and Ulitko, V.A. and Moskvin, A.S.},
issn = {15556190},
journal = {Physics of the Solid State},
number = {7},
pages = {1316--1321},
title = {{Modified Monte Carlo Method with the Heat Bath Algorithm for a Model Cuprate}},
volume = {67},
year = {2025}
}

@article{Miyatake1986,
doi = {10.1088/0022-3719/19/14/020},
year = {1986},
month = {may},
volume = {19},
number = {14},
pages = {2539},
author = {Y Miyatake and M Yamamoto and J J Kim and M Toyonaga and O Nagai},
title = {On the implementation of the 'heat bath' algorithms for Monte Carlo simulations of classical Heisenberg spin systems},
journal = {Journal of Physics C: Solid State Physics},
}

\end{document}